\documentclass[twocolumn]{autart}    

\usepackage{graphicx}          
\usepackage{multirow}
\usepackage{amsmath} 
\usepackage{amssymb}  
\usepackage{mathrsfs}
\usepackage{algorithmicx}
\usepackage{algpseudocode}
\usepackage[Algorithm,ruled]{algorithm}
\usepackage{nicefrac}

\begin{document}

\begin{frontmatter}

\title{Volterra Kernel Identification using Regularized Orthonormal Basis Functions\thanksref{footnoteinfo}} 

\thanks[footnoteinfo]{Corresponding author Jeremy G. Stoddard.}

\author[Stod]{Jeremy G. Stoddard}\ead{jeremy.stoddard@uon.edu.au},    
\author[Stod]{James S. Welsh}\ead{james.welsh@newcastle.edu.au}               

\address[Stod]{School of Electrical Engineering and Computing, The University of Newcastle, Australia}  

\begin{keyword}                           
System identification; Nonlinear systems; Volterra series; Basis functions; Regularization.               
\end{keyword}                             

%
%

\begin{abstract}                          

The Volterra series is a powerful tool in modelling a broad range of nonlinear dynamic systems. However, due to its nonparametric nature, the number of parameters in the series increases rapidly with memory length and series order, with the uncertainty in resulting model estimates increasing accordingly.  In this paper, we propose an identification method where the Volterra kernels are estimated indirectly through orthonormal basis function expansions, with regularization applied directly to the expansion coefficients to reduce variance in the final model estimate and provide access to useful models at previously unfeasible series orders. The higher dimensional kernel expansions are regularized using a method that allows smoothness and decay to be imposed on the entire hyper-surface. Numerical examples demonstrate improved Volterra series estimation up to the 4th order using the regularized basis function method.

\end{abstract}

\end{frontmatter}

%
%

\section{Introduction}

In the field of system identification, data-driven modelling of nonlinear systems poses unique challenges, which can not be completely addressed by the already well established linear identification theory.  One of the more significant issues in nonlinear identification is choosing a model structure from the vast array of possible model classes, which can require significant prior knowledge on the system.

The Volterra series provides a nonparametric representation for a broad range of nonlinear systems, and unlike parametric models, requires only limited prior knowledge to perform the series estimation.  While the theoretical advantages are clear, practical estimation of Volterra series models is a challenging task~\cite{Cheng2017}. The (truncated) series can be seen as a high dimensional generalization of the linear Finite Impulse Response (FIR), and much like FIR models, longer memory lengths are required for more accurate modelling. To compound this issue, the Volterra series must also be extended to kernels of higher dimension to capture the nonlinear behaviour of the underlying system.  This results in a large number of parameters for estimation, with a corresponding high variance in the model estimates obtained in the presence of measurement noise and finite data records.

Whilst large numbers of parameters are a distinct disadvantage for nonparametric models, several techniques have been developed to address this for FIR models in the linear case, and some of these methods have been extended to the nonlinear setting for low variance Volterra series estimation. The so called `Bayesian regularization' approach,  introduced in \cite{Pillonetto2010}, is one such method that shows promise in a Volterra series context. In the linear case, the variance of parameter estimates is reduced by imposing some degree of smoothness and exponential decay on the estimated impulse response, at the price of a small bias. Recently it was shown that the same properties can also be imposed on impulse responses of higher dimension \cite{Birpoutsoukis2017}. Orthonormal basis function modelling \cite{Heuberger2005} is another applicable technique, since the Volterra series can be expressed in terms of basis function expansions \cite{Rugh1980}. However, placement of the basis function poles is more difficult to optimize in multi-dimensional kernels than in the linear case \cite{Campello2004},\cite{Rosa2007}.

The contribution of this paper is to propose the direct regularization of basis function expansion coefficients, in order to reduce the variance of model estimates and also provide access to higher memory lengths and series orders. While the concept of regularized expansion coefficients has been explored for linear systems \cite{Chen2015}, its potential in the nonlinear setting is yet to be shown. This paper motivates the use of regularization on multi-dimensional basis function expansions, and provides a novel framework for the separable optimization of hyperparameters in the case of Laguerre and Kautz basis functions. Numerical results show the proposed method performing better than existing methods for low orders, while also identifying models at previously unfeasible series orders.

The paper is organized as follows. Section \ref{Volt} provides an overview of the Volterra identification problem, while Section \ref{OBF} introduces the required background on orthnormal basis function modelling.  Section \ref{Reg} gives details on the regularization approach taken in this paper, and a separable optimization method is developed in Section \ref{Opt} for pole selection in Laguerre and Kautz basis functions. The new identification methods are assessed in Section \ref{Num} through Monte Carlo simulations on several nonlinear systems. Finally, some conclusions are presented in Section \ref{Concl}.

%
%

\section{Volterra Kernel Identification}
\label{Volt}

Any causal, time-invariant and fading-memory nonlinear system can be well approximated using a truncated discrete-time Volterra series representation \cite{Boyd1985}. The series consists of a sum of Volterra kernels, where each kernel acts on products of lagged input values. The resulting model is linear-in-the-parameters, and can be estimated in a least squares framework, but the large numbers of parameters in the model can make the estimation quite computationally intensive.

\subsection{Volterra Model Description}

\begin{figure}[t]
\centering
\includegraphics[width = 0.45\textwidth]{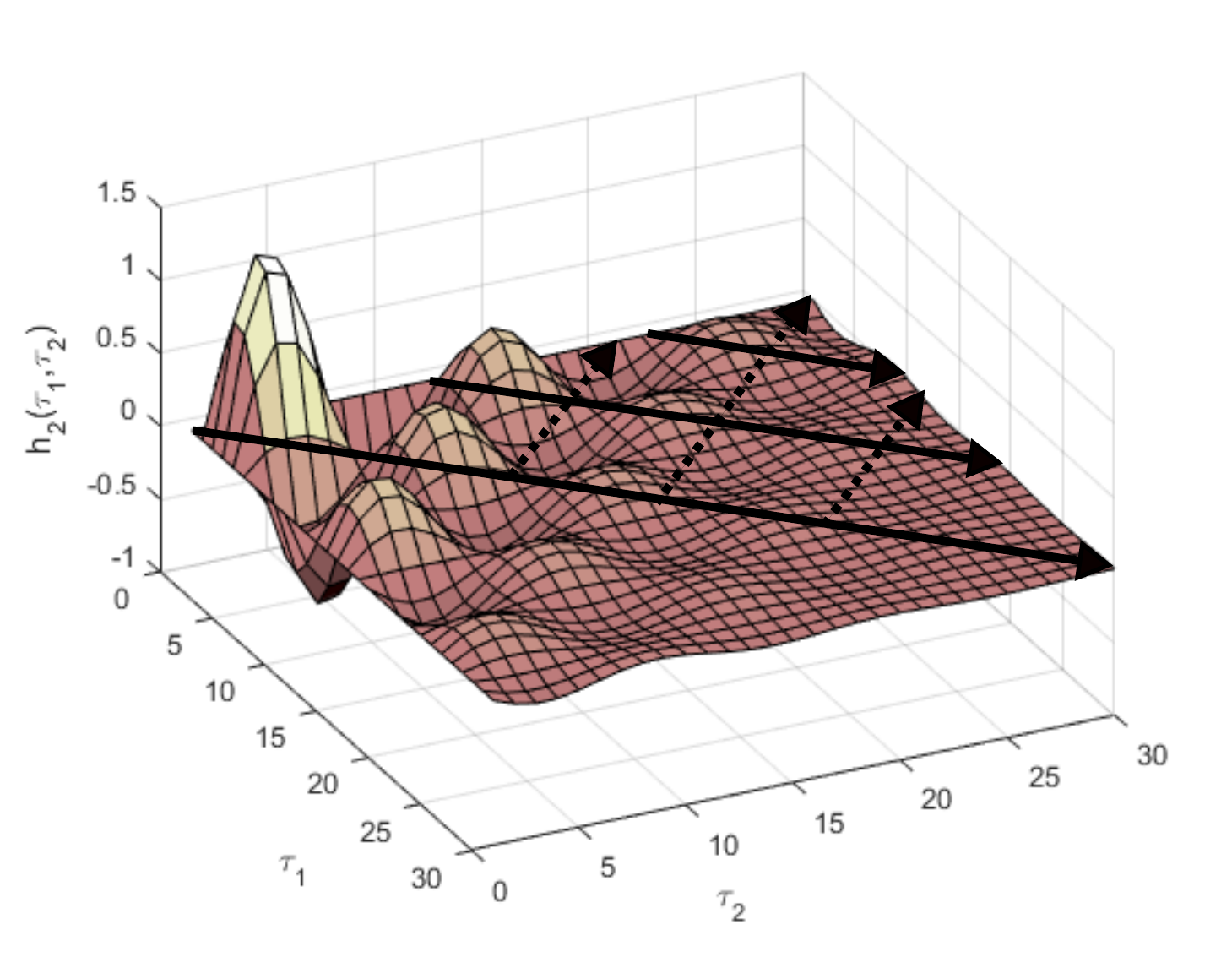}
\caption{Second order Volterra kernel of a resonant Wiener system, and the two perpendicular regularizing directions, (1,1) (solid) and (-1,1) (dotted)}
\label{RegDir}
\end{figure}

For an input series $u$ and noise-free output $y^0$, we consider the Volterra series model,
\begin{equation}
\label{volterra}
y^0(k) = \sum_{m=1}^{M} \Bigg[ \sum_{\tau_1=0}^{n_m - 1} \hdots \sum_{\tau_m=0}^{n_m-1} h_m(\tau_1,\hdots,\tau_m) \prod_{\tau = \tau_1}^{\tau_m} u(k-\tau) \Bigg],
\end{equation}  
where $m$ is the dimension of the kernels, $M$ is the maximum degree, $h_m(\tau_1,\hdots,\tau_m)$ is the $m$'th Volterra kernel, $n_m$ is the memory length of $h_m$, and $\tau_j$ is the $j$'th lag variable for the kernel. For $m>1$, the kernels can be viewed as (hyper)surfaces, such as the second order example of a resonant Wiener system shown in Figure \ref{RegDir}. 


\subsection{Least Squares Identification}
\label{LSident}

In this paper we assume that the input to the Volterra system is deterministic, and that white Gaussian measurement noise, $e$, is added directly at the output such that the measured output, $y$, is given by 
$$y(k) = y^0(k) + e(k),$$ 
where 
$$e(k) \sim \mathcal{N}(0,\sigma^2),$$
and $y^0$ is the noise-free Volterra series output from (\ref{volterra}). 

The kernel coefficients can be expressed as parameters in a least squares problem formulation, 
\begin{equation}
\label{LS}
Y_N = \phi_N^T \theta + E,
\end{equation}
where N is the number of measurements, $Y_N$ and $E$ are the vectors of output measurements and measurement noise respectively, with $\phi_N$ the regressor matrix corresponding to the vector of kernel coefficients, $\theta = [h_1^T, \bar{h}_2^T, \hdots \bar{h}_M^T]^T$. 

For higher-dimensions, symmetry is enforced in the kernels to ensure a unique representation \cite{Schetzen1980}. The \emph{unique} kernel coefficients from $h_m(\tau_1,\hdots, \tau_m)$ are taken and vectorized as $\bar{h}_m$ before being placed in the parameter vector, where the order of vectorization will determine the form of the regressor matrix.  

Under the assumed noise conditions, the Maximum Likelihood (ML) estimate of $\theta$ is given by the least squares analytic solution,
\begin{equation}
\label{LSsol}
\hat{\theta}_{LS} = ( \phi_N  \phi_N^T)^{-1} \phi_N Y_N.
\end{equation}

%
%

\section{Orthonormal Basis Function Representations}
\label{OBF}

We consider two particular basis function sets, which form a subset of the `Generalized Orthonormal Basis Functions' (GOBFs) \cite{Heuberger2005}. They are the Laguerre Basis Functions (LBFs) and 2-parameter Kautz Basis Functions (KBFs).

\subsection{Laguerre Basis Functions}

LBFs are formed as a first order realization of the GOBFs, and are parameterized by a single real pole.  In the $z$ domain, the functions take the form \cite{Heuberger2005}
\begin{equation}
\label{LBFdef}
F_i(z) = \frac{\sqrt{1-|a|^2}}{z-a} \bigg( \frac{1-az}{z-a} \bigg) ^{i-1}, \; \; \; a \in (-1,1).
\end{equation}
The Laguerre functions have a simple structure, but the absence of complex poles in the basis yields non-compact models for oscillatory systems \cite{Wahlberg1991}.  

\subsection{Kautz Basis Functions}
\label{Kautz}

KBFs are generated from second-order filters, such that complex pole pairs can be included to better model an oscillatory response. A practical parameterization \cite{Wahlberg1994} of 2-parameter Kautz functions is given as
\begin{equation}
\begin{split}
\label{KBFdef}
&F_{2i-1} = \frac{\sqrt{1-c^2}(z-b)}{z^2+b(c-1)z-c} \bigg( \frac{-cz^2 + b(c-1)z + 1}{z^2 + b(c-1)z - c} \bigg)^{i-1}, \\
&F_{2i} = \frac{\sqrt{(1-c^2)(1-b^2)}}{z^2+b(c-1)z-c} \bigg( \frac{-cz^2 + b(c-1)z + 1}{z^2 + b(c-1)z - c} \bigg)^{i-1},
\end{split}
\end{equation}
where $b,c \in (-1,1)$. 

\subsection{Basis Function Expansions of Volterra Models} 

Applying orthonormal basis functions to a Volterra series model is a concept referred to as the Volterra/Wiener approach \cite{Rugh1980}. For a kernel, $h_m$, as defined in (\ref{volterra}), the basis function expansion can be expressed as
\begin{equation}
\begin{split}
\label{OBF2Kernel}
h_m&(\tau_1,\hdots,\tau_m) \\ &= \sum_{i_1=1}^{B_m} \hdots \sum_{i_m=1}^{B_m} \alpha_m(i_1,\hdots,i_m) \prod_{j=1}^{m} f_{m,i_j}(\tau_j),
\end{split}
\end{equation}
where $f_{m,l}$ is the impulse response corresponding to the $l$'th basis function of the $m$'th kernel's basis, $B_m$ is the number of basis functions in the basis, and $\alpha_m(\cdot)$ is the set of expansion coefficients. Equations (\ref{volterra}) and (\ref{OBF2Kernel}) can be combined to restructure the Volterra model, i.e.
\begin{equation}
\label{OBFvolterra}
y(k) = \sum_{m=1}^{M} \Bigg[ \sum_{i_1=1}^{B_m} \hdots \sum_{i_m=1}^{B_m} \alpha_m(i_1,\hdots,i_m) \prod_{j=1}^{m} u^f_{m,i_j}(k) \Bigg],
\end{equation}
where $u^f_{m,l}$ is the input, $u$, filtered by the $l$'th basis function of the $m$'th kernel's basis. Note the similarity in structure between the models in (\ref{volterra}) and (\ref{OBFvolterra}), which motivates the treatment of the expansion coefficient sets $\alpha_m$ as `basis function kernels' in the domains of their corresponding bases. These new kernels can be much more compact than their time-domain counterparts, provided that the bases are carefully designed \cite{Campello2004},\cite{Rosa2007}.

Least squares identification of the $\alpha_m$ kernels is possible using the framework described in Section \ref{LSident}, with the regressor, $\phi_{f,N}$, now containing filtered input products. For the vectorised set $\alpha = [\alpha_1^T, \bar{\alpha}_2^T, \hdots, \bar{\alpha}_M^T]^T$, we have
\begin{equation}
\hat{\alpha}_{LS} = ( \phi_{f,N}  \phi_{f,N}^T)^{-1} \phi_{f,N} Y_N.
\label{LSbf}
\end{equation}

%
%

\section{Regularization of Kernel Estimates}
\label{Reg}

This section first presents an overview of the Bayesian regularization method from \cite{Pillonetto2010} and its extension to higher-dimensional kernels as developed in \cite{Birpoutsoukis2017}. We then introduce a novel application to basis function kernels.    

\subsection{Regularized Least Squares}
Considering the least squares problem outlined in Section \ref{LSident}, the regularized least squares problem is defined through the addition of a quadratic penalty on the parameter vector, $\theta$, such that the optimization problem is given by
\begin{equation}
\label{ReLS}
\hat{\theta}_{ReLS} = \text{arg } \underset{\theta}{\text{min}} \|Y_N - \phi_N^T \theta \|^2_2 + \sigma^2 \theta^T P^{-1} \theta,
\end{equation}
where $Y_N$ and $\phi_N$ are defined as in (\ref{LS}), and $P$ is a regularization penalty matrix. Taking a Bayesian perspective, $P$ can be interpreted as the prior covariance matrix of a Gaussian parameter vector (i.e. $\theta \sim \mathcal{N}(0,P)$). In the FIR model case, $P$ is designed to impose the prior knowledge of smoothness and exponential decay in impulse responses. There exists several tunable covariance structures to encode this prior information \cite{Pillonetto2014}. Here we will consider the Tuned/Correlated (TC) structure, where the $x,y$'th element of $P$ is given by,
\begin{equation}
\begin{split}
\label{TC}
&P(x,y) = \beta \lambda^{max(x,y)}, \\
&\beta \geq 0, \; 0 \leq \lambda < 1, \; \eta = [\beta,\lambda].
\end{split}
\end{equation}
The hyperparameters, $\eta$, are typically tuned via a marginal likelihood maximization \cite{Pillonetto2010}, given by 
\begin{equation}
\label{MAP}
\hat{\eta} = \text{arg } \underset{\eta}{\text{min }} Y_N^T \Sigma_Y^{-1} Y_N + \text{log det } \Sigma_Y,
\end{equation}
where $\Sigma_Y$ is the covariance matrix of $Y_N$ obtained from the joint distribution of $[\theta \; \; Y_N]^T$ \cite{Pillonetto2010}. The solution to (\ref{ReLS}) can then be computed as
\begin{equation}
\label{ReLSsol}
\hat{\theta}_{ReLS} = (P(\hat{\eta}) \phi_N \phi_N^T + \sigma^2 I)^{-1} P(\hat{\eta}) \phi_N Y_N.
\end{equation}
The noise variance, $\sigma^2$, is not typically known \textit{a priori}, and must also be estimated. In this paper, we place this variance in the hyperparameter vector, $\eta$, for tuning.


\subsection{Regularization for Higher-Dimensional Kernels}

An extension of the FIR regularization method to the Volterra series was developed in \cite{Birpoutsoukis2017}, which relies on the construction of separate penalty matrices for each kernel in the model. If the least squares problem is formulated as in (\ref{LS}), then the total penalty, $P$, in (\ref{ReLS}), is chosen to be a block diagonal matrix,
\begin{equation}
\label{KernelPenalty}
P = \begin{bmatrix}
       P_1 &  &  \mathbf{0} \\
        & \ddots & \\
        \mathbf{0} &  & P_M
     \end{bmatrix},
\end{equation}
where $P_m$ is the prior covariance of the $m$'th kernel~\cite{Birpoutsoukis2017}. 

While $P_1$, being one-dimensional, can still be constructed using (\ref{TC}), the covariance structures for multi-dimensional kernels must now impose smoothness and decay along the entire (hyper)surface. The approach suggested in \cite{Birpoutsoukis2017} is to consider $m$ perpendicular regularizing directions for the kernel $h_m$, where one direction is the vector $(1,\hdots,1)$. The regularizing directions for a second order resonant Wiener kernel are depicted in Figure \ref{RegDir} as an example.

The regularizing directions for $h_m$ form a rotated coordinate system which we will denote $(v_m^1, v_m^2, \hdots, v_m^m)$.  Using this coordinate system, standard covariance structures can be applied to generate a partial covariance for each regularizing direction. For the TC structure applied along direction $v_m^j$, the corresponding partial covariance is given by,
\begin{equation}
\label{TCext}
P_m^j(x,y) = (\lambda_m^j)^{max(x',y')},
\end{equation}
where $i'$ is the coordinate of $\bar{h}_m(i)$ on the $v_m^j$ axis. If $\bar{h}_m(i)$ is associated with lag values $\bar{\tau} = (\tau_1, \hdots, \tau_m)$, then $i' = \langle \bar{\tau} , v_m^j \rangle$. The total covariance matrix for the kernel is produced through element-by-element multiplication of the individual matrices \cite{Birpoutsoukis2017}, i.e.
\begin{equation}
\label{FinalPenalty}
P_m(x,y) = \beta_m \cdot P_m^1(x,y) \cdot \hdots \cdot P_m^m(x,y).
\end{equation}
where $\beta_m$ is a normalization hyperparameter.

Using the TC structure, there are now $m+1$ hyperparameters per kernel which contribute to the vector $\eta$ tuned in (\ref{MAP}). The large dimension of the search space and the non-convexity of the problem necessitate a global optimization method.

\subsection{Regularization for Basis Function Expansions}

The parallels between time-domain and basis function Volterra models have motivated the treatment of expansion coefficients as Volterra kernels in the domains of their bases.  A natural progression then would be to apply regularization to these new kernels in the same way it can be applied to standard Volterra kernels, i.e. impose smoothness and decay along the hyper-surfaces generated by the basis function expansions, using
\begin{equation}
\label{ReBF}
\hat{\alpha}_{ReLS} = \text{arg } \underset{\alpha}{\text{min}} \|Y_N - \phi_{f,N}^T \alpha \|^2_2 + \sigma^2 \alpha^T P^{-1} \alpha,
\end{equation}  
where $\alpha = [\alpha_1^T, \bar{\alpha}_2^T, \hdots, \bar{\alpha}_M^T]^T$ and $P$ is constructed and tuned as in the previous section. Indeed, the concept has already been explored for linear FIR modelling \cite{Chen2015}, where the TC covariance structure was applied in regularized estimation of Laguerre coefficients. 

\section{Separable Parameter Selection for Laguerre and Kautz Bases}
\label{Opt}

When regularization is applied to basis function expansions directly, the basis-generating parameters must be tuned as well. One approach suggested for the linear case \cite{Chen2015} is to place the basis-generating parameters with the existing hyperparameters in $\eta$, and tune them in the optimization of (\ref{MAP}). For nonlinear identification, however, we require a new basis for each kernel, and therefore several sets of generating parameters. To limit the search space of the non-convex optimization, we will develop a separable method for estimating the optimal Laguerre or Kautz basis function parameters prior to regularization. 

\subsection{Optimal Selection of Laguerre and Kautz Poles}
\label{BFPoleSelection}

When expanding Volterra kernels, the optimal generating parameters for Laguerre and Kautz bases were discussed in \cite{Campello2004} and \cite{Rosa2007} respectively, where optimality is defined by the minimization of error introduced from truncation of the basis function expansions to a finite length.

In the Laguerre case, the optimal pole, $a$, for the basis can be computed from an analytic function of the time domain kernel, $h_m(\tau_1, \hdots, \tau_m)$ \cite{Campello2004}. For the 2-parameter Kautz case, no analytic solution exists, but sub-optimal analytic estimates can be obtained by fixing $b$ and optimizing $c$. In this paper, by searching through an appropriately discretized space of $b \in (-1,1)$, we analytically compute the optimal $c=c^*$ for each $b$, and the corresponding cost, $J_m(b,c^*)$ from \cite{Rosa2007}. The parameter set which produced the global minimum for $J_m$ is the optimal choice for expansion of $h_m$. 

\subsection{Algorithm for Separable Optimization}
It is clear that optimal pole selection, as summarized in Section \ref{BFPoleSelection}, requires exact knowledge of the time-domain kernels, $h_m$. Since the kernels are the quantities we are required to estimate, we will instead employ a recursive approach to the problem of optimizing basis function parameters, which is summarized in Algorithm \ref{BFopt}. 
\begin{algorithm}[h]
\begin{algorithmic}[1]
\Require Data structures $Y_N$ and $\phi_N$ from (\ref{LS})
\State Obtain LS estimate of all kernels $\{h_m\}^{M}_{m=1}$ using  (\ref{LSsol}).
\State Compute optimal basis function parameter/s for each $h_m$ (outlined in Section \ref{BFPoleSelection}).
\State Using the parameter/s from Step 2, obtain a LS estimate of the basis function kernels $\alpha_m$ via (\ref{LSbf}).
\State Transform $\alpha_m$ kernels to time-domain kernels, $h_m$, using (\ref{OBF2Kernel}).
\State Repeat Steps 2-4 until basis function parameter/s converge within desired tolerance.
\end{algorithmic}
\caption{Optimization of LBF/KBF Parameters}
\label{BFopt}
\end{algorithm}
 
\begin{rem}
If the data length, $N$, is less than the number of parameters to be estimated in Step 1, then the user can instead initialize the algorithm by entering at Step 3 with an initial guess of the basis function parameters. If $N$ is also lower than the number of required basis function coefficients, then the algorithm cannot be used, and poles should be included as hyperparameters in (\ref{MAP}).
\end{rem}


%
%

\section{Numerical Simulation}
\label{Num}

A number of Monte Carlo simulation studies were conducted to evaluate the performance of the proposed regularized basis function method. The studies are performed on systems with disparate dynamics, to emphasise the different advantages of Kautz and Laguerre bases. The basis function methods also permit estimation of higher order kernels than their time-domain counterpart, due to the compact nature of basis function representations, and this is explored via third and fourth order systems.

\subsection{Simulation Settings}

All studies were performed on Wiener systems of the form,
\begin{equation}
y(k) = \sum_{m=1}^M \bigg( \frac{B_m q^{-1}}{A(q)}u(k)\bigg)^m +e(k),
\end{equation}   
where $q$ is the forward shift operator, and $e(k) \sim \mathcal{N}(0,\sigma^2)$ is Gaussian output noise added in each realization. Each term in the sum defines an $m$'th order Volterra kernel, and each kernel possesses the same denominator dynamics, with $B_m$ scaling the kernel's contribution to the output.

\begin{figure}[t]
\centering
\includegraphics[width = 0.5\textwidth]{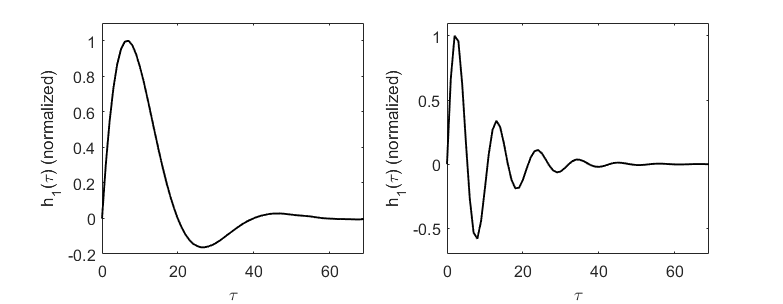}
\caption{Normalized impulse response of the linear block for Sys2a/Sys3/Sys4 (left) and Sys2b (right)}
\label{fig:Sys2a2bLinear}
\end{figure}

The simulation studies differ both in the resonance of the underlying linear filter, as well as the maximum kernel order, $M$. Two second order ($M=2$) systems were tested, with their dynamics given by
\begin{align*}
&\textbf{Sys2a: } A(q) = 1-1.8036q^{-1}+0.8338q^{-2} \\ 
&\textbf{Sys2b: } A(q) = 1-1.5q^{-1}+0.8125q^{-2}
\end{align*}
`Sys2a' is almost critically damped, while `Sys2b' has a lower damping ratio. The impulse response for each underlying linear filter can be seen in Figure \ref{fig:Sys2a2bLinear}. Studies were also performed on two higher order systems, `Sys3' ($M=3$) and `Sys4' ($M=4$), which use the same underlying filter as Sys2a, and approximately equal contributions to the output from each kernel.

For each system and method, Monte Carlo studies were performed at Signal to Noise Ratios (SNR) of 20dB and 5dB, with 100 system realizations per setting. The input, $u$, is constructed as a Gaussian distributed signal with unit variance, and the data length chosen to be $N=3412$; only 30\% longer than the minimum least squares requirement for Sys2a and Sys2b.

The methods proposed in this paper were evaluated alongside their unregularized counterparts, as well as a direct time-domain approach. The details of each method are:

\begin{enumerate}
\item \textbf{ReLS}: The regularized least squares method (from \cite{Birpoutsoukis2017}) in the time domain, using (\ref{KernelPenalty}), (\ref{TCext}) and (\ref{FinalPenalty}). 
\item \textbf{LBF}/\textbf{KBF}: Least squares estimation of LBF/KBF coefficients using (\ref{LSbf}). Generating parameters are pre-optimized using Algorithm \ref{BFopt}.
\item \textbf{ReLBF}/\textbf{ReKBF}: Regularized estimation of LBF/KBF coefficients using (\ref{ReBF}) and Algorithm \ref{BFopt}. 
\end{enumerate}

The time domain method was required to estimate each kernel up to a maximum lag $n_m=70$, while the basis function methods used a memory length of $B_m = 15$. For Sys3 and Sys4, ReLS becomes too computationally intensive to be feasible, hence only LBF and ReLBF methods were tested for these cases. For all regularized estimates, optimization of (\ref{MAP}) was performed via the $\text{\tt{GlobalSearch}}$ algorithm in MATLAB and using the $\text{\tt{fmincon}}$ function to find local minima.

\subsection{Results}

For each Monte Carlo study, the estimation errors are quantified with the validation error metric used in \cite{Birpoutsoukis2017}, calculated by applying an input of length 50,000 to both the true system and the estimated system and defining `normalised RMS error' as,
\begin{equation}
E_{NRMS} = \frac{rms(y_{val}-y_{mod})}{rms(y_{val})},
\end{equation}
where $y_{val}$ and $y_{mod}$ are the noise-less outputs of the true and estimated system respectively. 

\begin{figure}[b]
\centering
\includegraphics[width = 0.5\textwidth]{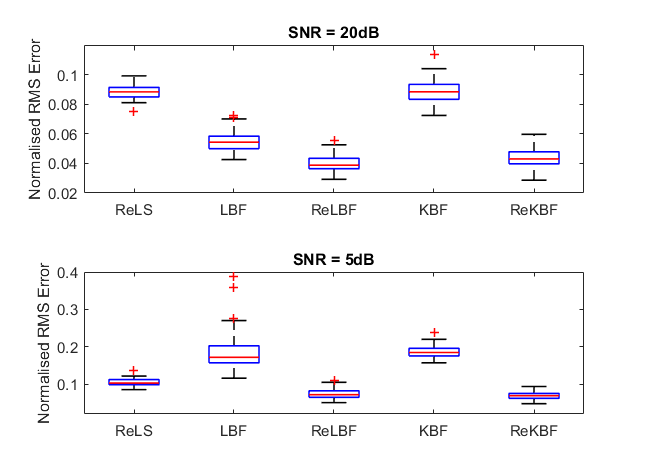}
\caption{Validation errors for Sys2a with SNR = 20dB (top) and 5dB (bottom)}
\label{Sys2a_Val}
\end{figure}

\begin{figure}[t]
\centering
\includegraphics[width = 0.5\textwidth]{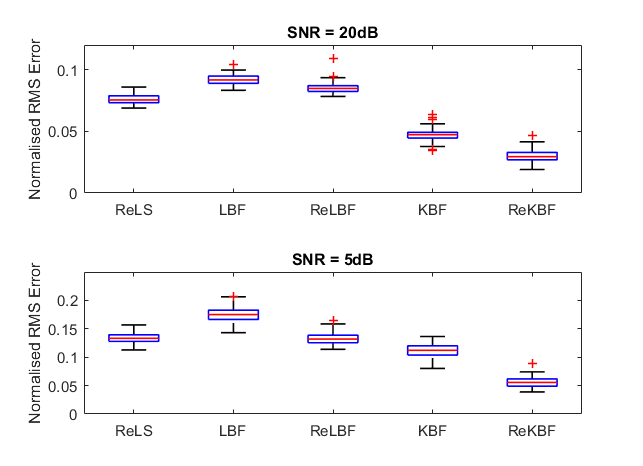}
\caption{Validation errors for Sys2b with SNR = 20dB (top) and 5dB (bottom)}
\label{Sys2b_Val}
\end{figure}

The second order results, presented as boxplots in Figures \ref{Sys2a_Val} and \ref{Sys2b_Val}, show significant improvements in output prediction using ReLBF and ReKBF. The advantage of Kautz functions for resonant systems is clear in the Sys2b (low damping) results, with ReKBF outperforming all other methods. For comparison, mean computation times are also provided in Table \ref{tab:MeanCompTimes_OBFs}. 

\begin{table}[b]
\centering
\caption{Mean computation times for 2nd order systems}
\label{tab:MeanCompTimes_OBFs}
\begin{tabular}{l|lllll}
Method & ReLS  & LBF & KBF  & ReLBF & ReKBF \\ \hline
Time (s) & 375.8 & 9.7 & 37.2 & 14.9  & 43.5 
\end{tabular}
\end{table}

\begin{figure}[t]
\centering
\includegraphics[width = 0.48\textwidth]{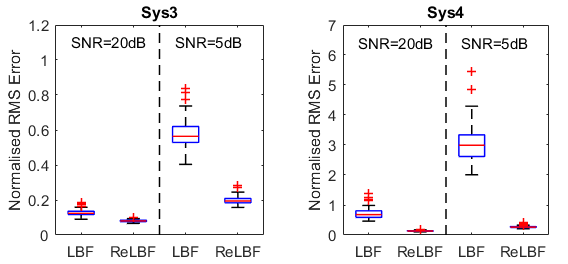}
\caption{Validation errors for Sys3 (left) and Sys4 (right)}
\label{Sys3and4_Val}
\end{figure}

The validation errors for Sys3 and Sys4 are given in Figure \ref{Sys3and4_Val}, which highlight the benefit of the regularized methods in lowering model error to tolerable levels.  At these orders, the newly proposed method could obtain estimates with reasonable accuracy and computation time using a standard computer architecture, which was not possible using exisiting methods. 

%
%

\section{Conclusion}
\label{Concl}

This paper makes a novel proposal to combine the techniques of basis function modelling and regularization for the purpose of Volterra series estimation, and provides the theoretical motivation for imposing standard covariance structures on basis function kernels. To reduce the complexity of the regularization cost function, a separable optimization method was designed for pre-selecting Laguerre/Kautz parameters for each kernel. The performance of the proposed estimation method has been compared against unregularized estimates and estimates directly in the time-domain, with improved output prediction in all test cases. Furthermore, the proposed methods allowed access to computationally feasible and accurate estimates at higher series orders than previously possible, even for low data length and high noise settings.

%
%

\vspace{-2mm}

\bibliographystyle{plain}        

\end{document}